\documentclass[preprint,showpacs,showkeys,preprintnumbers,amsmath,amssymb]{revtex4-1}
\usepackage[utf8]{inputenc}

\usepackage{graphicx,subfigure}
\usepackage{psfrag}
\usepackage{amsmath}
\usepackage{bbm}
\usepackage{dsfont}
\usepackage{braket}

\newcommand{\defgr}{\mathrel{\mathop:\!\!=}}

\begin{document}

\preprint{TUD-ITP-TQO/04-2010-V100303}

\title{Volumes of conditioned bipartite state spaces}

\author{Simon Milz}
     \affiliation{Institut f\"{u}r Theoretische Physik,
Technische Universit\"at Dresden, D-01062 Dresden, Germany}
\author{Walter T. Strunz}
   \affiliation{Institut f\"{u}r Theoretische Physik,
Technische Universit\"at Dresden, D-01062 Dresden, Germany}

\date{\today}

\begin{abstract}
We analyse the metric properties of \textit{conditioned} quantum state spaces $\mathcal{M}^{(n\times
m)}_{\eta}$. These spaces are the convex sets of $nm \times nm$ density matrices that, when partially
traced over $m$ degrees of freedom, respectively yield the given $n\times n$ density matrix $\eta$. For
the case $n=2$, the volume of $\mathcal{M}^{(2\times m)}_{\eta}$ equipped with the Hilbert-Schmidt
measure is a simple polynomial of the radius of $\eta$ in the Bloch-Ball. Remarkably, the probability
$p_{\mathrm{sep}}^{(2\times m)}(\eta)$ to
find a separable state in $\mathcal{M}^{(2\times m)}_{\eta}$ is independent of $\eta$ (except for
$\eta$ pure). Both these results are proven analytically for the case of the family of $4\times 4$
$X$-states, and thoroughly numerically investigated for the general case. The important implications
of these results for the clarification of open problems in quantum theory are pointed out
and discussed.
\end{abstract}

\pacs{03.67.-a, 42.50.Dv, 89.70.+c}
\keywords{Qubits, Qutrits, density matrix, Hilbert-Schmidt metric, entanglement, separability, quantum correlation,
Monte Carlo, numerical integration}

\maketitle

\section{Introduction}
\label{sec::Intro}
Open quantum system states $\eta$ are reduced system states $\eta=\rho_S=\mathrm{Tr}_R(\rho_{S+R})$ of some
total state $\rho_{S+R}$ of a system $S$ and its environment $R$, where $\mathrm{Tr}_R$ denotes the partial trace over the degrees of freedom of the environment. Open quantum system dynamics refers to the
time evolution $\eta \rightarrow \eta(t)$ determined through the unitary evolution of system and
environment \cite{alicki_quantum_1987}: $\eta(t)=\mathrm{Tr}_R\left(\mathrm{U}_{S+R}(t)\rho_{S+R}\mathrm{U}^{\dagger}_{S+R}(t)\right)$.
A crucial issue that is widely discussed (see for example
\cite{pechukas_reduced_1994,alicki_comment_1995,pechukas_pechukas_1995,
rodriguez-rosario_linear_2010}) is how to
map the state $\eta$ of the open system $S$ at some initial time to a total state of $S+R$
(formalized through the so-called {\it assignment map} $\pi$). In
the literature this assignment is always considered to be linear and most results are obtained on
the assumption that $\pi$ maps $\eta$ on a product with a fixed state of the environment, i.e.
$\pi\left(\eta\right) = \eta \otimes \rho_R$.

While the
mathematical properties of $\pi$ have been discussed in detail
\cite{rodriguez-rosario_linear_2010,jordan_assumptions_2006,masillo_remarks_2011}, only little is
known
\cite{jordan_dynamics_2004} about its image, i.e. the set of total states $\rho$ of the closed
system $S+R$ that are compatible with a given reduced state $\eta$ of the open system $S$. A
thorough investigation of these spaces is therefore necessary to obtain a more complete picture of
the properties of assignment maps, and thus, a more complete picture of open quantum
system dynamics.

Apart from their relevance for the description of open quantum dynamics, spaces of total states that are
conditioned to a given reduced state constitute lower-dimensional sections of the total state
space. An analysis of these sections might hence shed light on the properties of the total space. As
only little is known about general quantum dynamical state spaces (see e.g. \cite{kimura_bloch_2003}
for a discussion of the state space of a qubit and
\cite{zyczkowski_hilbertschmidt_2003,sommers_bures_2003} for the Hilbert-Schmidt and the Bures
volume of general state spaces) an investigation of a new kind of sections of these spaces might lead
the way to a solution of long-standing problems concerning quantum dynamical state spaces.

In their seminal work \cite{zyczkowski_volume_1998}, \.{Z}yczkowski et al. raised the question of
the volume of separable states in the total state space of a bipartite system and emphasized that
its solution is of both philosophical and experimental interest. Ever since, this problem has
been tackled for different measures both numerically and analytically. Analytical results are
at hand only for certain lower-dimensional sections of the total state space \cite{slater_exact_2000}.
For the general problem only conjectures based on extensive numerical
research exist \cite{slater_concise_2013}. The conjecture that is of most importance for this
paper is the belief that $\mathcal{P}_{\mathrm{sep}}^{(2\times 2)}$, the a priori
Hilbert-Schmidt-probability for a two-qubit state to be separable, is equal to $\frac{8}{33}$
\cite{slater_dyson_2007}. In spite of the existence of these analytical and numerical results,
a more general geometric picture of state correlations is highly desirable. Our results on
conditional state spaces presented here may help to shed light on some long-standing open problems
concerning geometrical considerations of state spaces.

Our paper is structured as follows: In section \ref{sec::StaSpa} and \ref{sec::HS-measure} the general framework of bipartite
systems is introduced. A possible parametrisation of these systems that will be used throughout
this paper, is presented. The state spaces are equipped with the
Hilbert-Schmidt measure as the measure for which all metric results will be derived. Section
\ref{sec::CoupledQubit} introduces coupled qubit systems, which are the lowest-dimensional possible
bipartite systems and therefore allow for a feasible numerical treatment. The main results of this
paper are to be found in sections \ref{sec::X-states}, \ref{Sec::Vol} and \ref{sec::CondProb}. In the
first of these, analytical results
for the Hilbert-Schmidt volume of spaces of conditioned $X$-states and the probability to find a
separable state in these spaces are derived, while the latter two constitute a numerical
investigation
of the metric properties of general coupled qubit-systems. A conclusion and a discussion of the implications
of the results are given in section \ref{sec::dis}.

\section{State spaces of bipartite quantum systems}
\label{sec::StaSpa}
An $N \times N$ density matrix $\rho$ is a bounded linear operator acting on the Hilbert space
$\mathcal{H}_N$ (i.e. $\rho \in \mathcal{B}(\mathcal{H}_N)$) that satisfies the following three
conditions:
\begin{enumerate}
 \item $\rho^{\dagger} = \rho$
\item $\mathrm{Tr}\rho = 1$
\item $\rho$ is positive semi-definite.
\end{enumerate}
The convex set of all $N\times N$ density matrices is denoted by $\mathcal{M}^{(N)}\subset
\mathcal{B}(\mathcal{H}_N)$. Because of the unit trace and the hermiticity of density matrices, the
number of free real parameters of $\rho$ is equal to $N^2-1$. The demand for positivity restricts the
domain of these parameters.

If $N=n\times m$ is not prime, $\rho \in \mathcal{B}(\mathcal{H}_{n\times m})$ can be considered as
a bipartite state consisting of an $n$-dimensional system $S$ coupled to an
$m$-dimensional system $R$. A natural parametrisation of an $nm \times nm$ density matrix $\rho \in
\mathcal{M}^{(n\times m)}$ makes use of the traceless generators of the special unitary group
$SU(nm)$ \cite{jordan_mapping_2006} (Einstein summation convention implied):
\begin{equation}
 \label{eqn::Para}
\rho = \frac{1}{nm}\left(\mathds{1}_{mn} + a_i A_i \otimes \mathds{1}_m + b_j \mathds{1}_n
\otimes B_j + c_{kl} A_k \otimes B_l\right)\, ,
\end{equation}
where $A_i$ and $B_j$ are the generators of the groups $SU(n)$ and $SU(m)$, respectively,
$a_i,b_j,c_{kl} \in \mathbb{R}$ and
$\mathds{1}_{mn}$ is the $nm\times nm$ identity matrix. $A_i$ and $B_j$ are chosen to satisfy the
standard
orthonormality relations 
\begin{equation}
 \label{eqn::orthogonality}
\mathrm{Tr}(A_kA_l)=2\delta_{kl} \quad \mathrm{and} \quad \mathrm{Tr}(B_kB_l)=2\delta_{kl}\, .
\end{equation}

The parametrisation given by equation \eqref{eqn::Para} is not the only one in use. In the
literature parametrisations that use the Cholesky decomposition
\cite{daboul_conditions_1967,slater_radial_2010} or the Euler angles of the elements of the group
$SU(N)$ \cite{tilma_generalized_2002,byrd_differential_1998} can also be found. Which parametrisation to employ
depends strongly on the problem to be solved. In the context of bipartite systems,
parametrisation \eqref{eqn::Para} is advantageous, as it directly implements the unity trace and
the hermiticity of $\rho$, and the states $\rho_S$ and $\rho_R$ of the systems $S$ and $R$ can be
inferred directly: 
\begin{equation}
 \label{eqn::SandR}
\rho_S = \mathrm{Tr}_{R} \rho = \frac{1}{n}\left(\mathds{1}_n + a_iA_i\right) \quad \mathrm{and}
\quad \rho_R = \mathrm{Tr}_{S} \rho = \frac{1}{m}\left(\mathds{1}_m + b_jB_j\right)\, ,
\end{equation}
where $\mathrm{Tr}_R$ and $\mathrm{Tr}_S$ denote the partial traces over the degrees of freedom of
the systems $R$ and $S$ respectively. In the following, the state $\rho \in
\mathcal{M}^{(n\times m)}$ will be called the total state, whereas $\eta=\rho_{\mathrm{S}} =
\mathrm{Tr}_{\mathrm{R}}\rho \in \mathcal{M}^{(n)}$ will be called the reduced state of $\rho$ to
emphasize the connection with open quantum system dynamics.

The vector $\vec \mu =
\left(a_1,\cdots,a_{n^2-1},b_1,\cdots,b_{m^2-1},c_{11},c_{12},\cdots,c_{21},\cdots,c_{n^2-1\, m^2-1}\right)^{\mathrm{T}}$
completely determines the state $\rho \in \mathcal{M}^{(n\times m)}$ and vice versa. The positivity of density
matrices restricts the possible vectors $\vec \mu$ to a proper subset $\Sigma^{(n\times m)}$ of
$\mathbb{R}^{(n^2m^2-1)}$.

In the framework of open system quantum mechanics, so called assignment maps are introduced
\cite{pechukas_reduced_1994}. These maps assign a compatible total state to each reduced
state of the open system. In the language of this article, an assignment map $\pi: \mathcal{M}^{(n)}
\rightarrow \mathcal{M}^{(n\times m)}$ is a map with the property
\begin{equation}
 \label{eqn::Assignment}
\eta \in \mathcal{M}^{(n)} \ \ \Rightarrow \ \ \pi(\eta)=\rho \in \mathcal{M}^{(n\times m)} \quad
\mathrm{and} \quad \mathrm{Tr}_{\mathrm{R}}(\rho) = \eta\, .
\end{equation}
For a given reduced state $\eta \in \mathcal{M}^{(n)}$, the total state $\rho \in
\mathcal{M}^{(n\times m)}$ with $\mathrm{Tr}_{\mathrm{R}}(\rho)=\eta$ is obviously not unique. 
In order to gain a better understanding of open quantum dynamics, it is
therefore necessary to investigate the spaces of total states $\rho \in \mathcal{M}^{(n\times m)}$ 
that are conditioned on a given reduced state $\eta$. These spaces will be denoted as
$\mathcal{M}^{(n \times m)}_{\eta}$: 
\begin{equation}
 \label{eqn::CondSpace}
\mathcal{M}^{(n\times m)}_{\eta} = \left\{ \rho \in \mathcal{M}^{(n\times m)}:
\mathrm{Tr}_{\mathrm{R}}\rho = \eta\in \mathcal{M}^{(n)}\right\}\, .
\end{equation}
Its corresponding subspace of $\mathbb{R}^{n^2m^2-1}$ will be denoted as $\Sigma^{(n\times
m)}_{\eta} \subset \Sigma^{(n\times m)}$.

Given that 
\begin{equation}
 \label{eqn::VereinRaeume}
\underset{\eta\in \mathcal{M}^{(n)}}{\bigcup} \mathcal{M}^{(n \times m)}_{\eta} =
\mathcal{M}^{(n \times m)}\, ,
\end{equation}
a thorough analysis of conditioned spaces will not only shed light on assignment maps, but also on
the properties of the total state space $\mathcal{M}^{(n\times m)}$. Before metric properties of
$\mathcal{M}^{(n \times m)}_{\eta}$ can be discussed, it is necessary to introduce the notion of 
measure in $\mathcal{M}^{(n\times m)}$. Then, for example its volume and the a priori probability 
to find a separable state when choosing a state $\rho \in \mathcal{M}^{(n \times m)}_{\eta}$ at 
random can be determined.

\section{The Hilbert-Schmidt measure}
\label{sec::HS-measure}
While the space of pure $nm$-dimensional states has a natural measure, the so-called Fubini-Study
measure \cite{study_kurzeste_1905}, there is no unique measure to choose in the space $\mathcal{M}^{(n\times
m)}$ of mixed states. As for the parametrisation, the choice of the employed measure depends on the
question that is to be answered. A comprehensive overview over a wide family of measures in
$\mathcal{M}^{(n\times m)}$ can be found in \cite{bengtsson_geometry_2008}.\newline \newline
One possibility to introduce the notion of distance that induces a measure in the space
$\mathcal{M}^{(n\times m)}$ makes use of the unitarily invariant Hilbert-Schmidt inner product
$\left(\cdot \, ,\cdot\right)_{\mathrm{HS}}:\ \mathcal{M}^{(n\times m)}\times \mathcal{M}^{(n\times
m)} \rightarrow \mathbb{C}$:
\begin{equation}
 \label{eqn::HS-Inner}
\rho^{\prime},\rho  \in \mathcal{M}^{(n\times m)}: \quad
\left(\rho^{\prime},\rho\right)_{\mathrm{HS}} \defgr
\mathrm{Tr}\left(\rho^{\prime}\rho^{\dagger}\right) = \mathrm{Tr}\left(\rho^{\prime}\rho\right)\, .
\end{equation}
Following this definition, the Hilbert-Schmidt distance
$d_{\mathrm{HS}}\left(\rho^{\prime},\rho\right)$ of two arbitrary density matrices
$\rho^{\prime},\rho \in \mathcal{M}^{(n\times m)}$ can be expressed as
\begin{equation}
\label{eqn::HS-dist}
d_{\mathrm{HS}}\left(\rho^{\prime},\rho\right) =
\sqrt{\mathrm{Tr}\left[\left(\rho-\rho^{\prime}\right)^2\right]}\, .
\end{equation}
The Hilbert-Schmidt distance induces a flat metric in $\Sigma^{(n\times m)}$ because of the
tracelessness of the generators of the group $SU(n\times m)$ and the orthonormality relations
\eqref{eqn::orthogonality}:
\begin{align}
d_{\mathrm{HS}}\left(\rho^{\prime},\rho\right) &=
\frac{\sqrt{2}}{nm}\sqrt{\sum_{i=1}^{n\times m}\left(a_i^{\prime}-a_i\right)^2+\sum_{j=1}^{n\times
m}\left(b_j^{\prime}-b_j\right)^2+ \sum_{k,l=1}^{n\times m}\left(c_{kl}^{\,
\prime}-c_{kl}\right)^2} \notag\\
 \label{eqn::FlatMet}
&= \frac{\sqrt{2}}{nm}d_{\mathrm{euclid}}\left({\vec \mu}^{\, \prime},\vec \mu\right)
\end{align}
Up to an insignificant constant (which could be set equal to one by a change of the normalization of
\eqref{eqn::Para}), the mapping 
\begin{equation}
 \label{eqn::MappingSpaces}
\left(\mathcal{M}^{(n\times m)},d_{\mathrm{HS}}\right) \ \ \rightarrow \ \
\left(\Sigma^{(n\times m)},d_{\mathrm{euclid}}\right)
\end{equation}
is bijective and isometric. Therefore, any metric results about $\mathcal{M}^{(n\times m)}$
equipped with the Hilbert-Schmidt distance directly give the corresponding metric result in
$\Sigma^{(n\times m)}$ equipped with the flat euclidian metric, and vice versa. This, of course, is
also true for any subspaces of $\mathcal{M}^{(n\times m)}$ and $\Sigma^{(n\times m)}$ respectively,
in particular for the spaces  $\mathcal{M}^{(n\times m)}_{\eta}$ and $\Sigma^{(n\times m)}_{\eta}$.

The Hilbert-Schmidt volume $V_{\mathrm{HS}}^{(n\times m)}$ of the space $\mathcal{M}^{(n\times m)}$
has been calculated by \.Zyczkowski and Sommers in \cite{zyczkowski_hilbertschmidt_2003}:
\begin{equation}
 \label{eqn::HS-Volume}
\displaystyle
V_{\mathrm{HS}}^{(n\times m)}= \sqrt{nm}(2\pi)^{nm(nm-1)/2} \ \frac{\prod_{k=1}^{n\times m}
\Gamma(k)}{\Gamma(n^2m^2)}\, ,
\end{equation}
where $\Gamma(k)$ is the Gamma function of $k$. The derivation of \eqref{eqn::HS-Volume} makes use of the fact 
that any density matrix $\rho \in
\mathcal{M}^{(n\times m)}$ can be represented as $\rho = \mathrm{U}\Lambda \mathrm{U}^{\dagger}$,
where $\Lambda = \mathrm{diag}\left(\lambda_1,\lambda_2,\cdots,\lambda_{nm}\right)$ is a positive
diagonal matrix with $\mathrm{Tr}\Lambda = 1$ and $\mathrm{U} \in U(n m)$ is a unitary $nm\times nm$ matrix. As the Hilbert-Schmidt distance $d_{\mathrm{HS}}$ is unitarily invariant,
its corresponding volume element $\mathrm{d}V_{\mathrm{HS}}$ can be written as a product measure 
\begin{equation}
 \label{eqn::ProdMeas}
\mathrm{d}V_{\mathrm{HS}} = \mathrm{d}\mu\left(\lambda_1,\cdots,\lambda_{nm}\right) \times
\mathrm{d}\nu_{\mathrm{Haar}}\, ,
\end{equation}
where $\mathrm{d}\mu\left(\lambda_1,\cdots,\lambda_{nm}\right)$ is a measure on the space of
positive diagonal $nm\times nm$-matrices $\Lambda$ with $\mathrm{Tr}\Lambda = 1$, i.e. the
$(nm-1)$-simplex, and $\mathrm{d}\nu_{\mathrm{Haar}}$ is a measure on the space of unitary $nm\times
nm$ matrices that is induced by the Haar-measure on $U(nm)$. Both these measures can be expressed
analytically and the Hilbert-Schmidt volume of $\mathcal{M}^{(n\times m)}$ can be calculated
without resorting to the particular parametrisation \eqref{eqn::Para}. \newline \newline
For conditioned spaces, the property that the Hilbert-Schmidt measure is of product form fails to
apply. It is still true that any density matrix $\rho \in \mathcal{M}^{(n\times m)}_{\eta}$ can be
represented as $\mathrm{U}\Lambda \mathrm{U}^{\dagger}$, but for a given diagonal matrix $\Lambda$,
only certain matrices $\mathrm{U} \in U(nm)$ lead to a density matrix $\rho \in
\mathcal{M}^{(n\times m)}_{\eta}$, while most $\mathrm{U}$ result in a state $\rho^{\prime}
\notin \mathcal{M}^{(n\times m)}_{\eta}$. The volume of the total space
$\mathcal{M}^{(n\times m)}$ can be obtained by integrating over the whole space of unitary
matrices \footnote{In fact, the integral in \cite{zyczkowski_hilbertschmidt_2003} is not evaluated
over the total group $U(nm)$, but merely over the flag manifold $U(nm)/U(1)^{nm}$. This does however not
change the argument.} independently of the entries of the matrix $\Lambda$. In the case of
conditioned spaces, however, this independence no longer exists. Therefore, the considerations which
led to
the result \eqref{eqn::HS-Volume} cannot be used in order to find the Hilbert-Schmidt
volume $V^{(n\times m)}_{\mathrm{HS}}(\eta)$ of
the conditioned spaces $\mathcal{M}^{(n\times m)}_{\eta}$. \newline \newline
While the Hilbert-Schmidt volume of $\mathcal{M}^{(n\times m)}$ has been calculated, there only
exist conjectures for the a priori probabilities $\mathcal{P}_{\mathrm{sep}}^{(2\times 2)}$ (see
\cite{slater_concise_2013} and references therein) and
$\mathcal{P}_{\mathrm{sep}}^{(2\times 3)}$ \cite{slater_priori_2003} to find a separable state in
$\mathcal{M}^{(2\times 2)}$ (the state space of two coupled qubits) and
$\mathcal{M}^{(2\times 3)}$ (the state space of a qubit coupled to a qutrit) equipped with the Hilbert-Schmidt measure. As for the calculation of the
volume of conditioned state spaces, the problem of finding general analytical results for
$\mathcal{P}_{\mathrm{sep}}^{(n\times m)}$ is related to the fact that the Hilbert-Schmidt measure
in the
space of separable states is not of product form. Moreover, there are no unambiguous criteria for
the distinction between separable and entangled states beyond the $2\times 3$ case
\cite{horodecki_separability_1996,peres_separability_1996-1}. The same holds of course for
the corresponding a priori probabilities $p_{\mathrm{sep}}^{(n\times m)}(\eta)$ to find a
separable state in the respective conditioned spaces $\mathcal{M}^{(n\times m)}_{\eta}$.\newline
\newline
The calculation of $V_{\mathrm{HS}}^{(n\times m)}(\eta)$ as well as the investigation of the a
priori probabilities $p_{\mathrm{sep}}^{(n\times m)}(\eta)$ are important in the context of open
quantum dynamics, but they also shed further light on the structure and the properties of the total state
space $\mathcal{M}^{(n\times m)}$. As the measures involved are complicated and not explicitly
known, it is unlikely that the techniques used in \cite{zyczkowski_hilbertschmidt_2003} and
\cite{sommers_bures_2003} can be employed in order to solve these problems. It proves
fruitful, however, to exploit the isometry of the metric spaces $\left(\mathcal{M}^{(n\times
m)}_{\eta},d_{\mathrm{HS}}\right)$ and $\left(\Sigma^{(n\times m)}_{\eta},d_{\mathrm{euclid}}\right)$, i. e.
make use of the particular parametrisation \eqref{eqn::Para}, in order to find both numerical
results and analytical conjectures for the Hilbert-Schmidt volume of $\mathcal{M}^{(n\times
m)}_{\eta}$ and the a priori probabilities $p_{\mathrm{sep}}^{(n\times m)}(\eta)$. 

\section{Coupled qubit-systems}
\label{sec::CoupledQubit}
The only quantum dynamical state space whose structure is completely known is the
three-dimensional one-qubit state space $\mathcal{M}^{(2)}$ (see for example
\cite{kimura_bloch_2003} for a thorough discussion of its properties). Expressed in the
parametrisation \eqref{eqn::Para}, adapted to a monopartite system, any qubit-state can be written as
\begin{equation}
 \rho_{\mathrm{Qubit}} = \frac{1}{2}\left(\mathds{1}_2 + a_i \sigma_i\right)\, ,
\end{equation}
where the operators $\sigma_i$ are the well-known Pauli-matrices. The constraint of positive
semi-definiteness restricts the so-called Bloch-vector $\vec r =
\left(a_1,a_2,a_3\right)^{\mathrm{T}}$ to a solid ball (the Bloch-ball) of radius $|\vec r\, | \leq
1$.
The pure qubit-states make up the surface of this ball, the mixed states lie in the interior and the
completely mixed state $\rho_* = \frac{1}{2}\mathds{1}$ is at the center of the ball.

The lowest-dimensional state-space of a bipartite system is the space $\mathcal{M}^{(2\times 2)}$,
i.e. the state space of two coupled qubits. Systems of coupled qubits play an important role in the
context of quantum computation (cf. \cite{nielsen_quantum_2000}) and the dimensions of
$\mathcal{M}^{(2\times 2)}$ and $\mathcal{M}^{(2\times 2)}_{\eta}$ are low enough to allow for
direct calculations in the coordinates of parametrisation \eqref{eqn::Para}. Therefore the
following investigations will be mainly restricted to the $2\times 2$ case. \newline \newline
Any two-qubit-state $\rho_{2\mathrm{Qubits}} \in \mathcal{M}^{(2\times 2)}$ can be written as 
\begin{equation}
 \label{eqn::2Qubits}
\rho_{2\mathrm{Qubits}} = \frac{1}{4}\left(\mathds{1}_4 + a_i\sigma_i\otimes \mathds{1}_2 +
b_j\mathds{1}_2\otimes \tau_j + c_{kl} \sigma_k \otimes \sigma_l\right)\, ,
\end{equation}
where $\sigma_i$ and $\tau_j$ are the Pauli-matrices of the qubits, respectively. The reduced state
$\eta = \mathrm{Tr}_{\mathrm{R}}\left(\rho_{2\mathrm{Qubits}}\right)$ is completely defined by the vector $\vec
a = \left(a_1,a_2,a_3\right)^{\mathrm{T}}$. All further considerations will be simplified by the
observation that both the Hilbert-Schmidt measure in the space $\mathcal{M}^{(2\times m)}_{\eta}$
and the
separability of a state $\rho_{2\mathrm{Qubits}} \in \mathcal{M}^{(2\times 2)}_{\eta}$ are invariant under
a transformation $\mathrm{W}\otimes \mathds{1}_m$, where $\mathrm{W}$ is an arbitrary special
unitary matrix $\in SU(2)$. As the group $SU(2)$ is the double cover of the group of
three-dimensional rotations, $SO(3)$ \cite{taylor_cover_1954}, $V_{\mathrm{HS}}^{(2\times
m)}(\eta)$ and $p_{\mathrm{sep}}^{(2\times m)}(\eta)$ only depend on the radius of $\eta$ in the
Bloch-ball. Accordingly, it is sufficient to calculate the volume $V_{\mathrm{HS}}^{(2\times
m)}(r)$ and the probability $p_{\mathrm{sep}}^{(2\times m)}(r)$ for a ray from the center of the
Bloch-ball to its surface, i.e. $r\in \left[0,1\right]$. In the following, this ray will be chosen
to be the ray from the center of the Bloch-ball to its north-pole, i.e. $a_1=a_2=0, a_3=r$. As the
calculations are too involved to be carried out analytically even for the two-qubit case, they will be 
conducted in a
first step for the seven-dimensional family of two-qubit $X$-states.

\section{Conditioned volume $V_{\mathrm{HS}}^{(X)}(r)$ and a priori
Hilbert-Schmidt separability probability $p_{\mathrm{sep}}^{(X)}(r)$ for $X$-states}
\label{sec::X-states}
\begin{figure}
 \centering
 \includegraphics[scale=0.3]{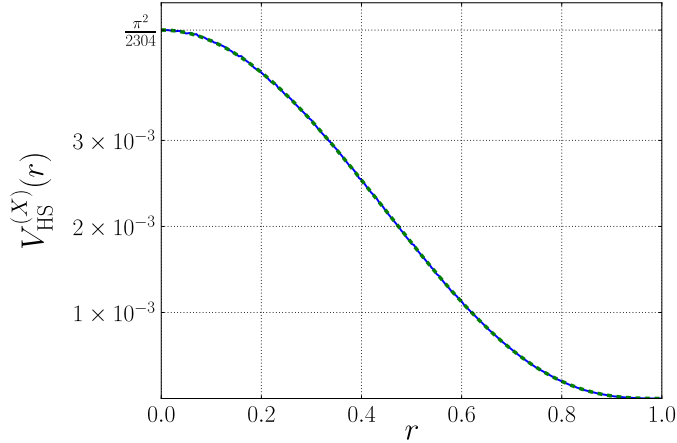}
 \caption{Hilbert-Schmidt volume $V^{(X)}_{\mathrm{HS}}(r)$ of the spaces of conditioned $X$-states.
Shown are the analytical solution
 \eqref{eqn::VolX-StatesCond} (green, dotted) and the numerical result of a Monte-Carlo
integration with
$10^7$ samples (blue).}
  \label{fig::VolXCond}
\end{figure}

A possible family of states in $\mathcal{M}^{(2\times 2)}$ that allows for analytical
conclusions, is the family of $X$-states. These are $4\times 4$ density matrices of the form
\begin{equation}
 \label{eqn::X-States}
\rho_X = \begin{pmatrix} \rho_{11}&0&0&\rho_{14} \\ 0&\rho_{22}&\rho_{23}&0 \\0&\rho_{32} &
\rho_{33} & 0 \\ \rho_{41} &0&0&\rho_{44} \end{pmatrix}\, .
\end{equation}
They have been introduced in \cite{maziero_classical_2009-1,vedral_entanglement_1998} as a
seven-dimensional family of states that contains maximally entangled pure states, as well as
separable states. Because of their simple form, it is possible to carry out various analytical
computations, like for example the calculation of the quantum discord of an
$X$-state \cite{ali_quantum_2010}. Note that the definition of $X$-states requires the choice of a
fixed basis. Here, the basis is chosen such that the Bloch-vector of the reduced state has a
$z$-component only (see below). $X$-states do not constitute a ''random'' subset of
$\mathcal{M}^{(2\times 2)}$, but possess an underlying symmetry \cite{rau_algebraic_2009}, which
might help to generalise the results found for $X$-states to arbitrary systems.\newline \newline
A comparison of \eqref{eqn::2Qubits} and \eqref{eqn::X-States} shows that any $X$-state can be
represented as 
\begin{align}
\rho_X &= \frac{1}{4} \left(\mathds{1} + a_3\sigma_3\otimes \mathds{1} + b_3 \mathds{1}\otimes
\tau_3 + c_{11}\sigma_1\otimes \tau_1 + c_{12}\sigma_1\otimes \tau_2 + c_{21}\sigma_2\otimes \tau_1
\right.\notag \\
\label{eqn::ParaX} 
&\phantom{=} \phantom{\frac{1}{4} (\mathds{1}\ \ }\left.+ c_{22}\sigma_2\otimes \tau_2 +
c_{33}\sigma_3\otimes \tau_3\right)\, .
\end{align}
As both $a_1$ and $a_2$ are equal to zero, the reduced states of $X$-states lie by construction on the ray from the center of the Bloch-ball to its north pole.
The eigenvalues of $\rho_X$ can be expressed analytically \cite{ali_quantum_2010}, and the two
eigenvalues that are of importance for the definiteness of $\rho_X$ lead to the conditions:
\begin{align}
 \label{eqn::X-Koord1}
\sqrt{\left(a_3+b_3\right)^2+\left(c_{11}-c_{22}\right)^2+\left(c_{12}+c_{21}\right)^2} \leq
\left(1+c_{33}\right) \, ,\\
\label{eqn::X-Koord2}
\sqrt{\left(a_3-b_3\right)^2+\left(c_{11}+c_{22}\right)^2+\left(c_{12}-c_{21}\right)^2} \leq
\left(1-c_{33}\right)\, .
\end{align}
The inequalities \eqref{eqn::X-Koord1} and \eqref{eqn::X-Koord2} define two hypersurfaces in
$\mathbb{R}^6$ that confine the space $\Sigma^{(X)}_{{a_3}}$ of $X$-states. The conditioned volume
$V_{\mathrm{HS}}^{(X)}(a_3)\equiv V_{\mathrm{HS}}^{(X)}(r)$ can be calculated by evaluating the
integral 
\begin{equation}
 \label{eqn::VolInteg}
\displaystyle V^{(X)}_{\mathrm{euclid}}(r) = \underbrace{\int \mathrm{d}c_{33}\int \mathrm{d} c_{11}
\int
\mathrm{d} c_{22} \int \mathrm{d} c_{12} \int \mathrm{d} c_{21} \int \mathrm{d} b_3}_{\text{Volume
defined by the inequalities (\ref{eqn::X-Koord1}) and (\ref{eqn::X-Koord2})}}
\end{equation}
and multiplying it by the appropriate factor, i.e. $V_{\mathrm{HS}}^{(X)}(r) =
\left(\frac{1}{2\sqrt{2}}\right)^6 V^{(X)}_{\mathrm{euclid}}(r)$. The explicit calculation is
carried out in appendix \ref{App::VolX}. It yields the surprisingly simple result 
\begin{equation}
 \label{eqn::VolX-StatesCond}
V_{\mathrm{HS}}^{(X)}(r) = \frac{\pi^2}{2304} \left(1-r^2\right)^3
\end{equation}
As a by-product of this formula, the Hilbert-Schmidt volume $V^{(X)}_{\mathrm{HS}}$ of the space of
$X$-states can be derived:
\begin{equation}
 \label{eqn::VolX-StatesTot}
\displaystyle
V^{(X)}_{\mathrm{HS}} = \int_0^1 \mathrm{d}r \, V_{\mathrm{HS}}^{(X)}(r) = \frac{\pi^2}{5040}
\end{equation}
In figure \ref{fig::VolXCond} we show the analytical curve $V^{(X)}_{\mathrm{HS}}(r)$ from \eqref{eqn::VolX-StatesCond} in
comparison with
numerical results obtained from a Monte-Carlo integration. \newline

For the special case of $X$-states, the partial transpose with respect to the second qubit only changes the signs of $c_{12}$ and $c_{22}$. 
The PPT-criterion (cf. \cite{horodecki_separability_1996,peres_separability_1996-1}) for the separability of a state together with 
inequalities \eqref{eqn::X-Koord1} and
\eqref{eqn::X-Koord2} then allows for a direct calculation of the volume of separable $X$-states, $V_{\mathrm{HS, sep}}^{(X)}(r)$.
The ratio determines
$p_{\mathrm{sep}}^{(X)}(r) = V_{\mathrm{HS,sep}}^{(X)}(r)/V_{\mathrm{HS}}^{(X)}(r)$, the probability to find a separable state in $\mathcal{M}_{r}^{(X)}$.
This calculation is carried out in appendix \ref{App::ProbX}. It yields: 
\begin{equation}
 \label{eqn::ProbX}
p_{\mathrm{sep}}^{(X)}(r) = \frac{2}{5}\, , \ r \in \left[0,1\right) \quad \mathrm{and} \quad
p_{\mathrm{sep}}^{(X)}(1) = 1\, .
\end{equation}
We also performed numerical Monte-Carlo calculations which confirmed these values. Most remarkably,
the probability to find a separable state in a conditioned state space $\mathcal{M}^{(X)}_r$ is
independent of the reduced state, i.e. independent of the radius $r$ for $r<1$ and jumps to one in a discontinuous 
way at $r=1$. The latter fact that $p_{\mathrm{sep}}^{(X)}(1) = 1$ is clear: a pure reduced state ($r=1$) can only
be realized by a product and thus, a separable total state.
 
\section{Conditioned volume $V_{\mathrm{HS}}^{(2\times m)}(r)$} 
\label{Sec::Vol}
While the eigenvalues of $X$-states can be easily expressed analytically, the eigenvalues of a
general two-qubit state could, in principle, be calculated. So far, however, a direct derivation of the volume
$V^{(2\times 2)}_{\mathrm{HS}}(r)$ from these expressions is beyond reach. For higher dimensional
cases, i.e. for a qubit coupled to an $m$-dimensional environment, not even the eigenvalues can be found
analytically. Accordingly, for $V_{\mathrm{HS}}^{(2\times m)}(r)$ and
$p_{\mathrm{sep}}^{(2\times m)}(r)$ only numerical results are provided here. The knowledge of
$V_{\mathrm{HS}}^{(X)}(r)$
and $V^{(2\times m)}_{\mathrm{HS}}$, however, allows for conjectures of the analytical expressions for
$V_{\mathrm{HS}}^{(2\times m)}(r)$ and $p_{\mathrm{sep}}^{(2\times m)}(r)$. \newline \newline
The Hilbert-Schmidt volume $V_{\mathrm{HS}}^{(2\times 2)}(r)$ is numerically estimated by a
Monte-Carlo integration. It can be readily derived that the range of each of the parameters $a_i$,
$b_j$ and $c_{kl}$ in the normalization of \eqref{eqn::2Qubits} is $\left[-1,1\right]$. Therefore,
a cube of edge length $d=2$ centred around the origin completely encompasses each conditioned
two-qubit space $\Sigma_r^{(2\times 2)}$ and its euclidian volume can be estimated by a simple
rejection sampling. Figure \ref{fig::VolXCond} shows the accuracy of this procedure for the
six-dimensional case of conditioned $X$-states, where we compare to analytical results. 
The full two-qubit problem is $12$-dimensional.
The higher dimension requires a larger number of samples in order to achieve similar accuracy.
For the general case of a qubit coupled to an $m$-dimensional environment, the
enormous number of samples necessary to obtain representative numerical results renders
rejection sampling methods useless.

\begin{figure}
 \centering
 \includegraphics[scale=0.3]{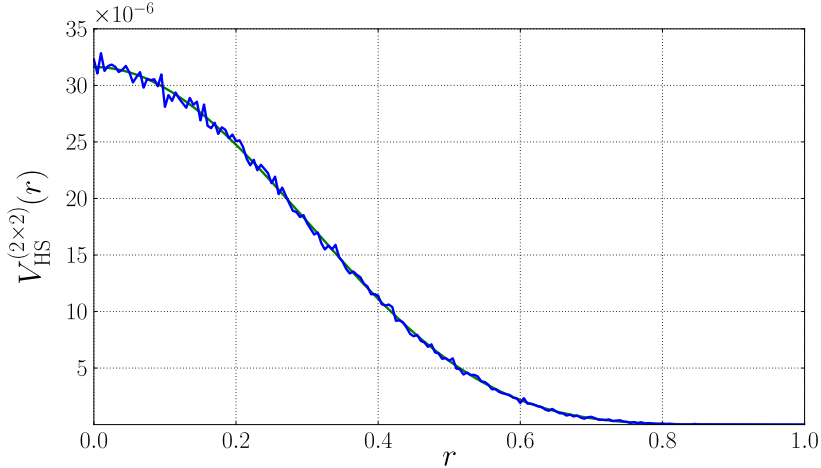}
 \caption{Hilbert-Schmidt volume $V^{(2\times 2)}_{\mathrm{HS}}(r)$ of the space of two coupled
qubits. Shown are conjecture
 \eqref{eqn::conjVol} (green) and the numerical result of a Monte-Carlo
 integration with $10^8$
 samples (blue).}
   \label{fig::Vol}
 \end{figure}

\begin{figure}
 \begin{center}
 \subfigure[$2\times 3$ case]{%
  \includegraphics[width=0.47\textwidth]{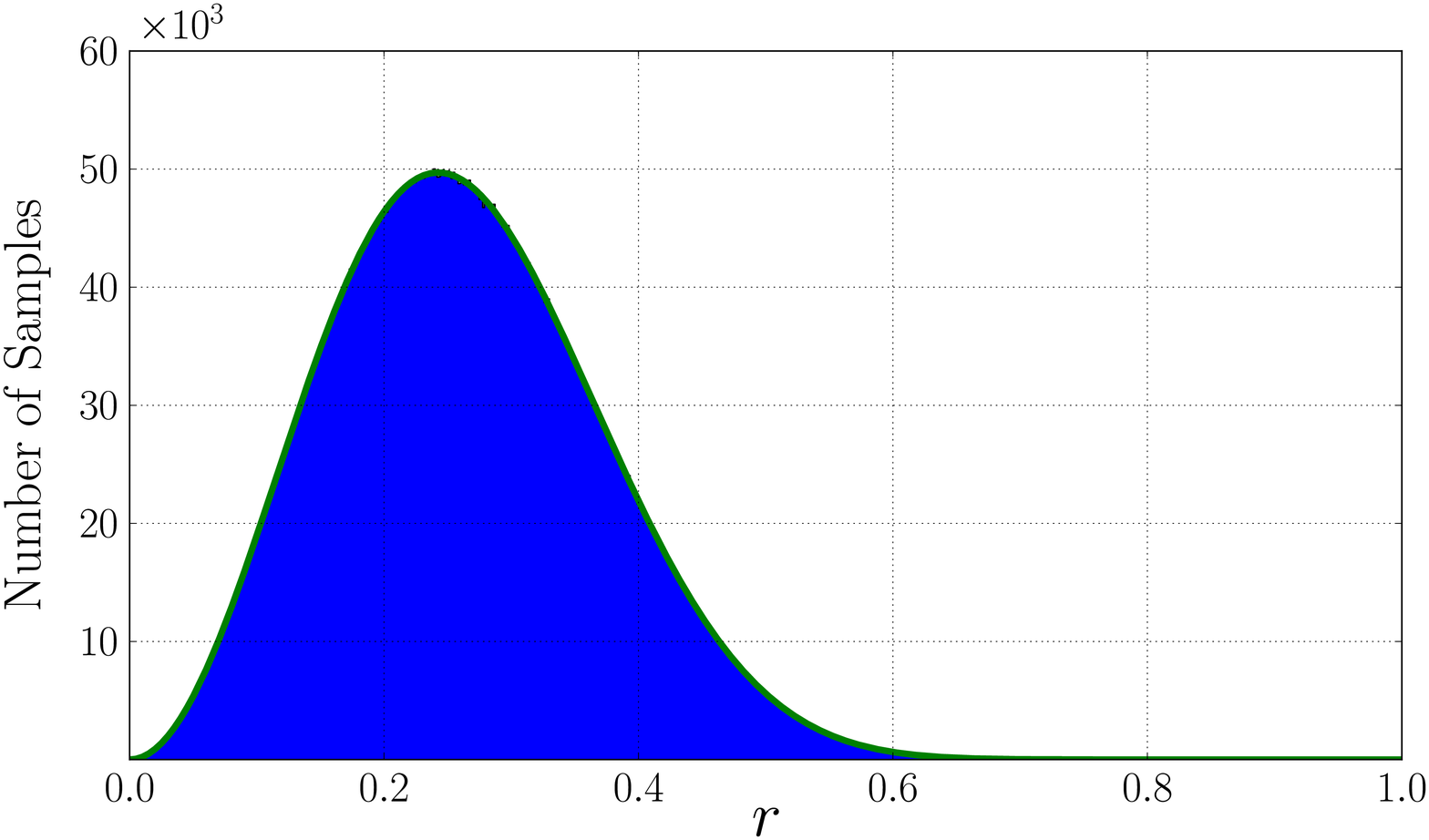}
  }%
 \hspace{0.02\textwidth}
  \subfigure[$2\times 4$ case]{%
  \includegraphics[width=0.47\textwidth]{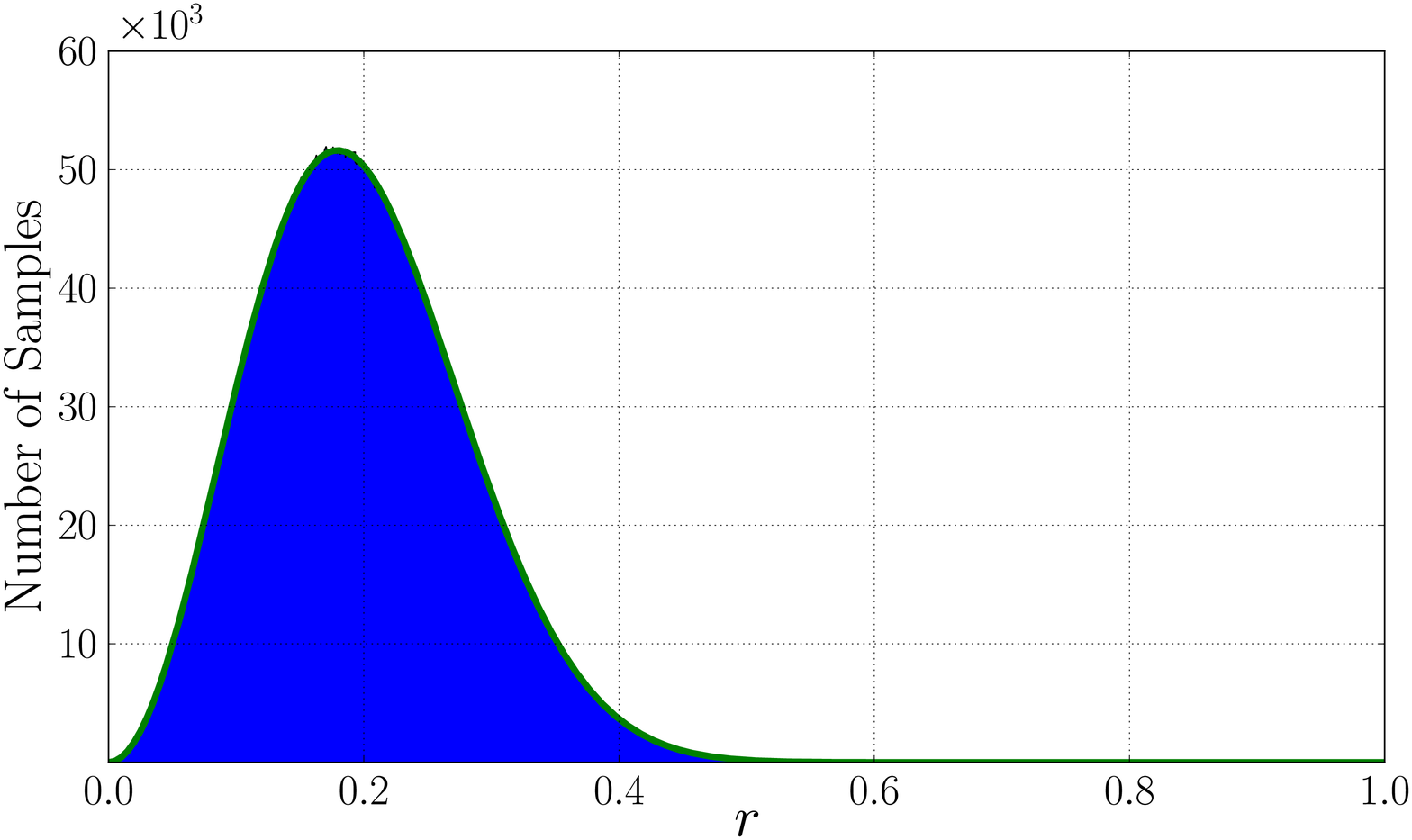}
  }
  \end{center}
\caption{Histograms of the Hilbert-Schmidt volume distribution as a function of the radius $r$ in
the Bloch-ball for the $2\times 3$ and the $2\times 4$ case
for $10^7$ samples, respectively. In blue the numerical data, in green the conjectured
envelopes.}
    \label{fig::Volgen}
 \end{figure}

A sampling method that goes without the rejection of sampling points makes use of the fact that
$N\times N$ density matrices can be sampled uniformly distributed according to the Hilbert-Schmidt
measure, by sampling pure $N^2$-dimensional states uniformly distributed according to the
Fubini-Study measure and partially tracing them over $N$ degrees of freedom (cf.
\cite{bengtsson_geometry_2008}). Applied to the case of $2m\times 2m$ density matrices, this means
that they can be sampled according to Hilbert-Schmidt measure, by sampling $4m^2$-dimensional pure
states according to the Fubini-Study measure (see e.g. \cite{bengtsson_geometry_2008}, chapter $7$,
for a description of how to sample pure states uniformly distributed according to the Fubini-Study
measure) and partially tracing these states over $2m$ degrees of freedom. In order to sample states $\rho \in \mathcal{M}^{(2\times m)}_{\eta}$ conditioned on a
given qubit state $\eta$ according to
Hilbert-Schmidt measure, it is sufficient to restrict the sampling of the pure states to the subset
of pure states that yield the given qubit state $\eta$ when partially traced over $2m^2$ degrees of
freedom. By construction, every sample then gives a valid $2m\times 2m$ density matrix, which
increases the accuracy of the results and makes it independent of $r$. This sampling method is hence well suited to
estimate $p_{\mathrm{sep}}^{(2\times m)}(r)$ and the course of $V_{\mathrm{HS}}^{(2\times m)}(r)$.
However, it does not yield any estimates of the absolute values of $V_{\mathrm{HS}}^{(2\times m)}(r)$
and, therefore, has to be combined with the results of the rejection sampling. \newline \newline
The result for $V^{(2\times 2)}_{\mathrm{HS}}(r)$ obtained from a Monte-Carlo integration with
$10^8$ samples is displayed in figure \ref{fig::Vol}.
The course of the numerical result resembles the analogous
analytical result for $V_{\mathrm{HS}}^{(X)}(r)$. For scaling reasons, the highest power of $r$ in
an analytical expression for $V_{\mathrm{HS}}^{(2\times 2)}(r)$ has to be $r^{12}$ which leads to
the conjecture that $V^{(2\times 2)}_{\mathrm{HS}}(r)$ is given by
\begin{equation}
\label{eqn::conjVol}
 V^{(2\times 2)}_{\mathrm{HS}}(r) = V_{\mathrm{HS}}^{(2\times 2)}(0) \left(1-r^2\right)^{6}\, .
\end{equation}
The fit of the conjectured curve to the numerical data is shown in figure \ref{fig::Vol}. Assuming
that equation \eqref{eqn::conjVol} is correct, the value of $V_{\mathrm{HS}}^{(2\times 2)}(0)$ can
be calculated by connecting $V_{\mathrm{HS}}^{(2\times 2)}(r)$ to the volume
$V_{\mathrm{HS}}^{(2\times 2)}$ of the total two-qubit state space:
\begin{equation}
 \label{eqn::V0}
V_{\mathrm{HS}}^{(2\times 2)} = 2^{-3} \cdot 4\pi \int_0^1\mathrm{d}r\, r^2
V_{\mathrm{HS}}^{(2\times 2)}(r)=\frac{2^{9} \pi}{45045}V_{\mathrm{HS}}^{(2\times 2)}(0)\, ,
\end{equation}
where the factor $2^{-3}$ is necessary to convert from the euclidean to the Hilbert-Schmidt
volume \footnote{Note that equation (\ref{eqn::V0}) is not a relation of volumes.
$V_{\mathrm{HS}}^{(2\times 2)}(0)$ is a $12$-dimensional volume, while $V_{\mathrm{HS}}^{(2\times
2)}$ is a $15$-dimensional one. They cannot be compared in a meaningful way. Therefore
(\ref{eqn::V0}) is merely an equation to calculate the
numerical value of $V_{\mathrm{HS}}^{(2\times 2)}(0)$}. From \eqref{eqn::HS-Volume} it can then
be deduced that
\begin{equation}
 \label{eqn::V0_2}
V_{\mathrm{HS}}^{(2\times 2)}(0) = \frac{45045}{2^{9} \pi} V_{\mathrm{HS}}^{(2\times 2)} \approx
3.16241
\times 10^{-5}\, .
\end{equation}
This value coincides perfectly with the analogous value found via the Monte-Carlo sampling:
\begin{equation}
 \label{eqn::NumV0}
V_{\mathrm{HS,num}}^{(2\times 2)}(0) = (3.16333 \pm 0.05713) \times 10^{-5}\, .
\end{equation}
The agreement of the conjectured formula and the numerical results, and the fact that
$V_{\mathrm{HS}}^{(2\times 2)}(r)$ seems to be described by a simple polynomial, suggest the
following generalization of \eqref{eqn::conjVol} for the $2\times m$ case:
\begin{equation}
 \label{eqn::conjVolgen}
V_{HS}^{(2\times m)}(r) = V_{HS}^{(2\times m)}(0)\left(1-r^2\right)^{2(m^2-1)}\, .
\end{equation}
The corresponding conjecture for $V_{HS}^{(2\times m)}(0)$ follows as in
\eqref{eqn::V0}:
\begin{equation}
 \label{eqn::V0gen}
V_{HS}^{(2\times m)}(0) = \sqrt{m}\cdot 2^{6m^2-m-\frac{23}{2}}\cdot \pi^{2m^2-m-\frac{3}{2}} \cdot
\frac{\prod_{k=1}^{2m} \Gamma(k) \cdot \Gamma\left(\frac{1}{2}+2m^2\right)}{\Gamma(4m^2)\cdot
\Gamma(-1+2m^2)}\, .
\end{equation}
It is rather difficult to investigate the validity of \eqref{eqn::conjVolgen} and
\eqref{eqn::V0gen} with a numerical procedure that involves the rejection of samples, as the
dimension of the corresponding state spaces grows rapidly and the huge number of required samples to
obtain meaningful results cannot be reached within an acceptable amount of time even for the case
$m=3$. \newline \newline
However, by employing the method described above (not relying on the rejection of
samples) at least \eqref{eqn::conjVolgen} can be verified numerically. This is done by sampling
states
uniformly distributed according to the Hilbert-Schmidt measure, and recording their radius in the
Bloch-ball, respectively. The resulting histogram then has an envelope that is described by a
function proportional to $4\pi r^2V_{HS}^{(2\times m)}(r)$, where $V_{HS}^{(2\times m)}(r)$
is the conjectured formula \eqref{eqn::conjVolgen}. It is necessary to multiply by the factor $4\pi
r^2$, which is the area of the respective spheres, in order to correctly describe the envelope of
the histograms as they display the number of states sampled for a given radius of the reduced
states. The histograms for the $2\times 3$ and the $2\times 4$ case are shown in figure
\ref{fig::Volgen}. They coincide perfectly with \eqref{eqn::conjVolgen}. \newline

\section{Conditioned a priori Hilbert-Schmidt separability probability $p_{\mathrm{sep}}^{(2\times
m)}(r)$}
\label{sec::CondProb}
\begin{figure}
 \centering
\includegraphics[scale=0.3]{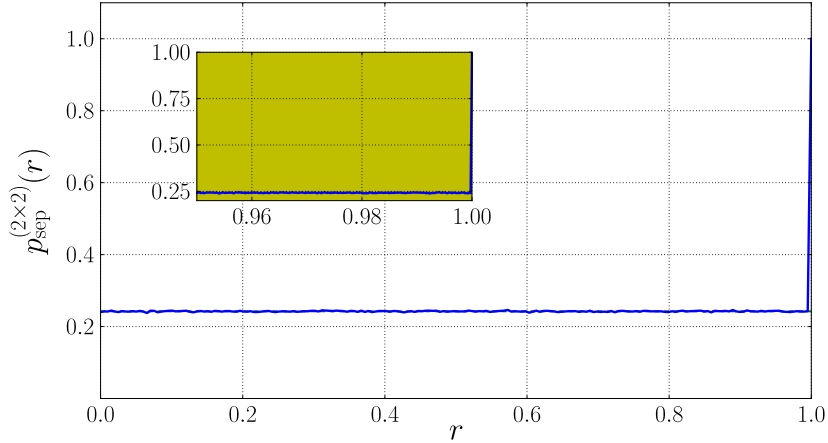}
\caption{Numerical results for the a priori Hilbert-Schmidt separability
probability $p_{\mathrm{sep}}^{(2\times 2)}(r)$ for $10^5$ samples. The small box shows the numerical results for $r\lessapprox 1$ in detail.}
   \label{fig::ProbSep2x2}
\end{figure}

Numerical results for the probability $p_{\mathrm{sep}}^{(2\times m)}(r)$ can easily be obtained
with high
accuracy by employing the sampling method that does not require the rejection of
samples. To this end, for each radius $r$, $k$ total states with the given radius are sampled, and,
at least for $m=2$ and $m=3$, the separability of each of these states is checked via the
PPT-criterion. The ratio of the number of separable states to the total number of sampled states
then gives an estimate for $p_{\mathrm{sep}}^{(2\times m)}(r)$. \newline \newline 
As it is not the absolute volume of the space of separable states that is to be estimated, but
rather the ratio of the two volumes $V_{\mathrm{HS,sep}}^{(2\times m)}$ and
$V_{\mathrm{HS}}^{(2\times m)}$, this sampling procedure in this case does not only give
qualitative
but also quantitative results. For the $2\times 2$ case they are displayed in figure
\ref{fig::ProbSep2x2}. \newline

The resemblance to the corresponding results for the $X$-states is striking:
our numerical evidence strongly suggests that
$p_{\mathrm{sep}}^{(2\times 2)}(r)$ is constant for $r\in \left[0,1\right)$ and jumps to
$1$ in a discontinuous way. As for $X$-states it is clear why $p_{\mathrm{sep}}^{(2\times 2)}(1)=1$.
The fact that this jump is discontinuous cannot yet be formally proven, but
the thorough numerical investigation of the region $r \lessapprox 1$, also shown in figure
\ref{fig::ProbSep2x2}, strongly suggests this conjecture. Obviously, from a knowledge of
$p_{\mathrm{sep}}^{(2\times 2)}(r)$, the value $\mathcal{P}_{\mathrm{sep}}^{(2\times 2)}$ of the
total state
space could be computed to be 
\begin{equation}
 \label{eqn::ProbSepQubitQubit}
\mathcal{P}_{\mathrm{sep}}^{(2\times 2)} = \int_0^1 \mathrm{d}r \, p_{\mathrm{sep}}^{(2\times
2)}(r)\, .
\end{equation}
The numerical results for $p_{\mathrm{sep}}^{(2\times 2)}(r)$ obtained above yield 
\begin{equation}
 \label{eqn::ProbSepQubitQubitNum}
\mathcal{P}_{\mathrm{sep,num}}^{(2\times 2)} = 0.24262 \pm 0.01340\, .
\end{equation}
Apart from a postulated but not yet formally proven formula \cite{slater_concise_2013}, there do
not exist
any analytical results for $\mathcal{P}_{\mathrm{sep}}^{(2\times 2)}$. However, based on extensive
numerical
research, a value of 
\begin{equation}
 \label{eqn::ProbSepQubitQubitConj}
\mathcal{P}_{\mathrm{sep}}^{(2\times 2)} = \frac{8}{33} \approx 0.24242
\end{equation}
has been conjectured (cf. \cite{slater_concise_2013} and references therein). The agreement between
\eqref{eqn::ProbSepQubitQubitNum} and \eqref{eqn::ProbSepQubitQubitConj} further supports the
conjecture of this value.

\begin{figure}
 \centering
\includegraphics[scale=0.3]{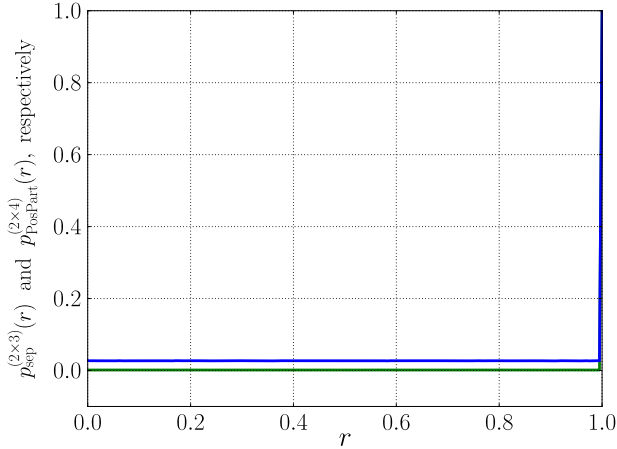}
\caption{Numerical results for the a priori Hilbert-Schmidt probabilities
$p_{\mathrm{sep}}^{(2\times 3)}(r)$ (blue) and
$p_{\mathrm{PosPart}}^{(2\times 4)}(r)$ (green) for $10^6$ samples. $p_{\mathrm{sep}}^{(2\times
3)}(r) \approx 0.0270$ and $p_{\mathrm{PosPart}}^{(2\times 4)}(r) \approx 0.0013 \, $ for $r\neq1$.}
   \label{fig::Sep2xm}
\end{figure}

The accuracy of the sample method without rejection even allows for an expansion of the numerical
investigation of $p_{\mathrm{sep}}^{(2\times 3)}(r)$ and the probability to find a state with
positive partial trace, $p_{\mathrm{PosPart}}^{(2\times 4)}(r)$. The respective results are shown in
figure \ref{fig::Sep2xm}.

They correspond qualitatively to the according results for $p_{\mathrm{sep}}^{(2\times 2)}(r)$.
Most remarkably, within numerical confidence,
$p_{\mathrm{sep}}^{(2\times 3)}(r)$ and $p_{\mathrm{PosPart}}^{(2\times 4)}(r)$ are
independent of $r$ -- except for the case $r=1$. For the $2\times 3$ case a numerical investigation
of $\mathcal{P}_{\mathrm{sep}}^{(2\times 3)}$ was conducted by Slater in \cite{slater_priori_2003}.
It
yielded: 
\begin{equation}
 \label{eqn::ProbSep2x3}
\mathcal{P}_{\mathrm{sep,\text{{\cite{slater_priori_2003}}}}}^{(2\times 3)} = 0.02631\, .
\end{equation}
From the numerical results above we find that
\begin{equation}
 \label{eqn::ProbSep2x3num}
\mathcal{P}_{\mathrm{sep,num}}^{(2\times 3)} = 0.02700 \pm 0.00016\, ,
\end{equation}
which is in good correspondence with \eqref{eqn::ProbSep2x3}. \newline \newline 
The independence of the
functions
$p_{\mathrm{sep}}^{(2\times 2)}(r)$, $p_{\mathrm{sep}}^{(2\times 3)}(r)$ and
$p_{\mathrm{PosPart}}^{(2\times 4)}(r)$ of $r$ leads to the conjecture that 
\begin{equation}
  \label{eqn::ConjProbSep}
 p_{\mathrm{sep}}^{(2\times m)}(r) = \mathcal{P}_{\mathrm{sep}}^{(2\times m)} \ \ \mathrm{for} \ \
r\in
\left[0,1\right) \ \ \mathrm{and} \ \ p_{\mathrm{sep}}^{(2\times m)}(1) = 1\, .
 \end{equation}
As beyond the $2\times 3$ case there is no simple criterion to decide whether or not a bipartite
state is separable, it is easier to test the more conservative conjecture
\begin{equation}
 \label{eqn::ConjPosPart}
p_{\mathrm{PosPart}}^{(2\times m)}(r) = \mathcal{P}^{(2\times m)}_{\mathrm{PosPart}} \ \
\mathrm{for} \ \
r\in
\left[0,1\right) \ \ \mathrm{and} \ \ p_{\mathrm{PosPart}}^{(2\times m)}(1) = 1\, ,
\end{equation}
where $\mathcal{P}^{(2\times m)}_{\mathrm{PosPart}}$ denotes the a priori Hilbert-Schmidt
probability for a state in $\mathcal{M}^{(2\times m)}$ with a positive partial transpose.

\section{Conclusion and discussion}
\label{sec::dis}
Although certain lower dimensional sections of the space $\mathcal{M}^{(2\times m)}$ have already
been studied analytically (see e.g. \cite{slater_exact_2000}), not much attention has been given to
the conditioned spaces $\mathcal{M}^{(2\times m)}_{\eta}$ yet. A thorough knowledge of their
properties is crucial for the understanding of assignment maps in the framework of open quantum
systems and might help to shed light on fundamental open questions in quantum state geometry.

In this work, the metric properties of the conditioned spaces $\mathcal{M}^{(2\times m)}_{\eta}$
equipped with the Hilbert-Schmidt measure have been investigated numerically and it turned out that
the Hilbert-Schmidt volume $V_{\mathrm{HS}}^{(2\times m)}(\eta)$ follows a simple polynomial of the radius
$r$ in the Bloch sphere of the reduced state $\eta$, while the probability to find a separable state in a
conditioned space $\mathcal{M}^{(2\times m)}_r$ is independent of $r$ -- except for the case $r=1$.
Both these results can be proven analytically for the case of the seven-dimensional family of
$X$-states.

Above all, the independence of $p_{\mathrm{sep}}^{(2\times m)}(r)$ of the radius $r$ we found 
is intriguing, as it, once analytically proven, opens new ways to study properties of the total
state space through these conditional cuts.
%

\begin{figure}
 \centering
\includegraphics[scale=0.3]{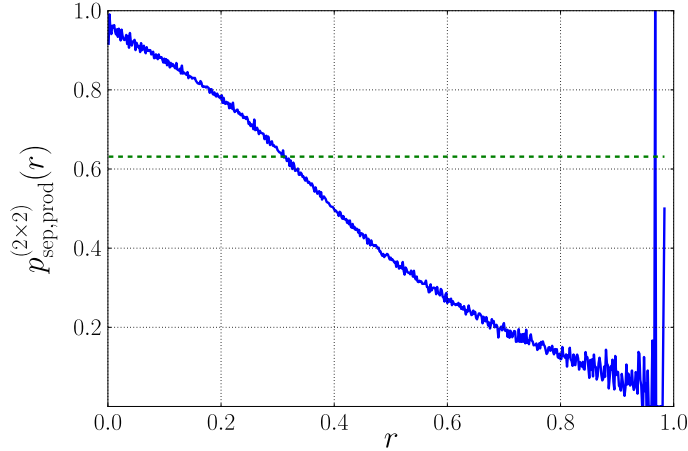}
\caption{Results for the a priori probability $p_{\mathrm{sep,prod}}^{(2\times 2)}(r)$ for the
product measure used in \cite{zyczkowski_volume_1998} obtained from a Monte-Carlo integration
with $10^7$ samples. Because of the vanishing volumes that are involved for $r\gtrapprox 0.9$
the relative error of the data diverges from this value on. In green, dotted, the numerical value of
$\mathcal{P}_{\mathrm{sep,prod}}^{(2\times 2)}$ for this measure is displayed.}
   \label{fig::naiveMeasure}
\end{figure}

It is important to point out that all these results and conjectures
only hold for the Hilbert-Schmidt measure. This particularity further singles out this measure
amongst all other unitarily invariant measures. The corresponding results for the example of the
product measure used in \cite{zyczkowski_volume_1998} are shown in figure
\ref{fig::naiveMeasure}. \newline \newline
One mean to find analytical expressions for all quantities investigated in this paper could be
the use of $X$-states. Obviously, there is a deep qualitative connection between the metric
properties of this seven-dimensional family of states, and the corresponding total space of states.
A thorough investigation of higher-dimensional $X$-states, i.e. in a first step $6\times 6$
$X$-states, might therefore further support the conjectures made for the general space
$\mathcal{M}^{(2\times m)}_r$. Furthermore, they might even be helpful from a quantitative point of
view. While the conjectured value for $\mathcal{P}_{\mathrm{sep}}^{(2\times 2)}$ of $\frac{8}{33}$
seems
plausible, the origins of this simple fraction still remain unclear. 
From formula \eqref{eqn::SepX} the volume of conditioned entangled $X$-states can be derived to be equal to $\frac{2}{15}\left(1-r^2\right)^3$. If $d_X$ denotes the number of free parameters for conditioned $X$-states, the denominator
$15$ is equal to $3\left(d_X-1\right)$. If $d_{(2\times 2)}$ denotes the corresponding number for
the full problem, it can easily be seen that the denominator $33$ of the conjectured value of
$\mathcal{P}_{\mathrm{sep}}^{(2\times 2)}$ is equal to $3\left(d_{(2\times 2)}-1\right)$. While this
is still
highly speculative, it nevertheless suggests that the analytical results for $X$-states might be
generalisable to the total $2\times 2$ and even higher-dimensional cases.

\section*{Acknowledgments}
It is a pleasure to thank Karol \.Zyczkowski for fruitful discussions and helpful advice.

\begin{appendix}
\section{Volume of the space $\mathcal{M}^{(X)}(r)$}
\label{App::VolX}
Semi-definiteness of $\rho_X$ yields two inequalities that define the seven-dimensional
subspace occupied by X-states in the full two-qubit parameter space:
\begin{eqnarray}
\label{eqn::ineq1}
 \sqrt{(a_3+b_3)^2+(c_{11}-c_{22})^2 + (c_{12}+c_{21})^2} \leq (1+c_{33})\, , \\
\label{eqn::ineq2}
\sqrt{(a_3-b_3)^2+(c_{11}+c_{22})^2 + (c_{12}-c_{21})^2} \leq (1-c_{33})\, . 
\end{eqnarray}
(\ref{eqn::ineq1}) and (\ref{eqn::ineq2}) set the limits for $c_{33}$: $c_{33} \in
\left[-1,1\right]$. Given these limits, (\ref{eqn::ineq1}) and (\ref{eqn::ineq2}) can be squared,
and one obtains:
\begin{eqnarray}
\label{eqn::ineq1trans}
 \left(a_3+b_3\right)^2+\left(c_{12}+c_{21}\right)^2+\left(c_{11}-c_{22}\right)^2-\left(1+c_{33}
\right)^2\leq 0 \\
\label{eqn::ineq2trans}
\left(a_3-b_3\right)^2+\left(c_{12}-c_{21}\right)^2+\left(c_{11}+c_{22}\right)^2-\left(1-c_{33}
\right)^2 \leq 0 
\end{eqnarray}
The task of calculating the volume defined by these two inequalities is remarkably simplified by the following coordinate transformation: 
\begin{alignat*}{2}
x&=\frac{1}{\sqrt{1-a_3^2}}\left(c_{22}+c_{11}\right)\, , \quad
&&X=\frac{1}{\sqrt{1-a_3^2}}\left(c_{21}+c_{12}\right)\, ,\\
y&=\frac{1}{\sqrt{1-a_3^2}}\left(c_{12}+c_{21}\right)\, , \quad &&Y=\frac{1}{\sqrt{1-a_3^2}}\left(c_{11}-c_{22}\right)\, ,
\\
z&=\frac{1}{1+a_3}\left(b_3+c_{33}\right) \, , \quad
&&Z=\frac{1}{1-a_3}\left(b_3-c_{33}\right)\, .
\end{alignat*}

This transformation is singular for $a_3=1$. However, for this value of $a_3$ the space $\mathcal{M}^{(X)}(a_3)$ is lower-dimensional, and its volume therefore zero. The Jacobian of this transformation is $\frac{1}{8}\left(1-a_3^2\right)^3$ and in the new coordinates \eqref{eqn::ineq1trans} and \eqref{eqn::ineq2trans} read as:
\begin{eqnarray}
\label{eqn::ineq1trans1}
x^2+y^2 &\leq& \frac{1}{1-a_3^2}\left[\left(1-c_{33}\right)^2 - \left(a_3-b_3\right)^2\right] \defgr r^2\, , \\
\label{eqn::ineq2trans1}
 X^2+Y^2&\leq& \frac{1}{1-a_3^2}\left[\left(1+c_{33}\right)^2 - \left(a_3+b_3\right)^2\right] \defgr R^2\, ,
\end{eqnarray}
with 
\begin{eqnarray*}
 r^2 &=& \frac{1}{1-a_3^2}\left[1-c_{33}-a_3+b_3\right]\left[1-c_{33}+a_3-b_3\right] = \left(1+Z\right)\left(1-z\right) \\
\mathrm{and} \quad R^2 &=& \frac{1}{1-a_3^2}\left[1+c_{33}-a_3-b_3\right]\left[1+c_{33}+a_3+b_3\right] = \left(1-Z\right)\left(1+z\right)\, .
\end{eqnarray*}
From \eqref{eqn::ineq1trans1} and \eqref{eqn::ineq2trans1} it follows that the integrals in the $x-y$-plane and the $X-Y$-plane 
will merely give the areas of circles of radius $r$ and $R$, respectively. As both $r^2$ and $R^2$ have to be positive, $z$ and $Z$ range from $-1$ to $1$. The euclidian volume $V_{\mathrm{euclid}}^{(X)}(a_3)$ of the space of conditioned $X$-states can then be calculated:
\begin{eqnarray}
\displaystyle
V_{\mathrm{euclid}}^{(X)}(a_3) &=& \frac{\pi^2}{8}\left(1-a_3^2\right)^3 \int_{-1}^1\mathrm{d}z\int_{-1}^1\mathrm{d}Z \underbrace{\left(1-Z^2\right)\left(1-z^2\right)}_{=r^2\cdot R^2} \notag \\
 \label{eqn::VolumeX}
&=& \frac{\pi^2}{8}\left(1-a_3^2\right)^3 \left[\int_{-1}^1\mathrm{d}z \left(1-z^2\right)\right]^2 = \frac{2}{9}\pi^2 \left(1-a_3^2\right)^3\, ,
\end{eqnarray}
which is valid for all $a_3 \in \left[0,1\right)$. From \eqref{eqn::VolumeX}, the result \eqref{eqn::VolX-StatesCond} for $V_{\mathrm{HS}}^{(X)}(a_3)$ follows directly.

\section{A priori probability $p_{sep}^{(X)}(r)$ to find a separable state in $\mathcal{M}^{(X)}_r$}
\label{App::ProbX}
In order for a two-qubit-state to be separable, its partial transpose with respect to one of the subsystems has to be positive semi-definite \cite{horodecki_separability_1996,peres_separability_1996-1}. For the special case of $X$-states, the partial transpose with respect to the second qubit merely changes the signs of $c_{12}$ and $c_{22}$. An $X$-state is hence separable iff it satisfies the two additional restrictions 
\begin{eqnarray}
 x^2 + y^2 &\leq& (1-Z)(1+z) = R^2\\
\mathrm{and} \quad X^2 + Y^2 &\leq& (1+Z)(1-z) = r^2\, .
\end{eqnarray}
Together with \eqref{eqn::ineq1trans1} and \eqref{eqn::ineq2trans1}, these two inequalities allow for a direct calculation of $V_{\mathrm{euclid,sep}}^{(X)}$, the euclidian volume of separable $X$-states: 
\begin{equation}
\displaystyle
V_{\mathrm{euclid,sep}}^{(X)} = \frac{\pi^2}{8}\left(1-a_3^2\right)^3 \int_{-1}^1 \mathrm{d}z\int_{-1}^1\mathrm{d}Z \ \min\left(r^2,R^2\right)^2\, ,
\end{equation}
where $\min\left(r^2,R^2\right)^2$ denotes the minimum of $\left\{r^2,R^2\right\}$. A short calculation yields that 
\begin{equation*}
 r^2 < R^2 \quad \Leftrightarrow \quad Z < z
\end{equation*}
and therefore 
\begin{equation}
 \label{eqn::CalcVolSep}
\displaystyle
V_{\mathrm{euclid,sep}}^{(X)} = \frac{\pi^2}{8}\left(1-a_3^2\right)^3  \int_{-1}^1 \mathrm{d}z \int_{-1}^z\mathrm{d}Z \ r^4(z,Z) + \int_{-1}^1 \mathrm{d}Z \int_{-1}^Z\mathrm{d}z \ R^4(z,Z). 
\end{equation}
It is easy to verify that \eqref{eqn::CalcVolSep} leads to the result
\begin{equation}
\label{eqn::SepX}
 V_{\mathrm{euclid,sep}}^{(X)} = \frac{4\pi^2}{45}\left(1-a_3^2\right)^3\, .
\end{equation}
Accordingly, for $r\equiv a_3 \in \left[0,1\right)$, the probability
$p^{(X)}_{\mathrm{sep}}(r)$ is independent of $r$ and equal to:
\begin{equation*}
p^{(X)}_{\mathrm{sep}}(r) = \frac{V^{(X)}_{\text{euclid,sep}}(r)}{V^{(X)}_{\mathrm{euclid}}(r)} = 2/5
\end{equation*}
For $r = 1$ all reduced states are pure, and therefore all the corresponding total states are separable. Hence: $p^{(X)}_{\mathrm{sep}}(1) = 1$.
\end{appendix}

\bibliography{Bib}
\bibliographystyle{plain}

\end{document}